\documentclass[prl,preprint,amsmath,amssymb]{revtex4}
\usepackage{graphicx}
\usepackage{color}
\usepackage{hyperref}
\usepackage[justification=centering]{caption}
\usepackage[sans]{dsfont}
\usepackage{subfig}

\begin{document}
\title{Effect of environment in Heisenberg $XYZ$ spin model }
\author{Indrajith.V.S}
\author{R. Sankaranarayanan}
\affiliation{Department of Physics, National Institute of Technology\\ Tiruchirapalli-620015, Tamil Nadu, India.}
\bigskip
\begin{abstract}
Quantum correlation of bipartite states (beyond entanglement) in presence of environment is studied for Heisenberg $XYZ$ spin system. It is shown that if the system is allowed to exchange energy with environment, the initial state evolves and settles down to uncorrelated state in asymptotic limit. We have also demonstrated that fidelity based measurement induced non-locality is a useful quantity in characterizing correlated quantum states.
\end{abstract}
\maketitle
\section{Introduction}
Quantum regime comprises of many interesting features that do not have analogy in the classical world. One such important feature is the non-local correlation. Non-locality refers to weird correlation between subsystems that make up a composite system \cite{epr}. In addition to its fundamental importance in understanding many body quantum states, the non-locality is a magical resource in various quantum information processing. Since a pure entangled state violates Bell inequality \cite{bel} - a test of non-locality, entanglement is considered as a manifestation of quantum correlation. Later Werner showed that there are unentangled mixed states that violate Bell inequality \cite{werner}. Since then, there is a demand for a bigger picture to capture all aspects of non-local correlation.

The formulation of discord using mutual information proved that all non-local correlated states need not be entangled \cite{discord}. This opened a new window for the study of non-local correlation beyond entanglement. Further development happened when Luo and Fuo proposed a new measure of correlation, namely Measurement Induced Non-locality (MIN) using von-Neumann projective measurement \cite{min}. Later on, other forms of MIN were studied as measures of correlation \cite{tmin,fmin} to resolve the so called local ancilla problem inherited with MIN \cite{piani}.

Further, unitary evolution of a closed quantum system is insufficient for realistic case, wherein the system frequently interacts with environment. Since such interaction significantly affects the non-local correlation of the system, it is worth studying the environmental intervention modelled by noisy quantum channels. An elegant way of investigating the influence of environment on quantum correlation is the operator-sum representation. Though we intuitively expect that the correlation and so the quantum signature of the system may be spoilt by the environment, it is important to study the dynamics of such process in detail to develop realistic information processing tasks. 
 
With this motivation, here we investigate quantum correlation of Heisenberg $XYZ$ spin system with magnetic field under the influence of environment modelled by Generalized Amplitude Damping (GAD) and hybrid channels \cite{corltn}. In the case of GAD channel, the system is permitted to exchange energy with the environment in the form of spontaneous emission and absorption processes. On the other hand, hybrid channel is a noisy environment inducing the operations namely, Bit-Flip, Phase-Flip and Bit-Phase-Flip with certain probabilities. The correlation between the spins in presence of the quantum channel is measured using MIN, Trace-MIN and Fidelity-MIN, along with concurrence - a measure of entanglement. In the case of energy exchange, we show that the system asymptotically settles down to uncorrelated state. Our detailed study on measuring correlation beyond entanglement reveals that Fidelity-MIN is more useful than its companion in characterizing bipartite quantum states.

 \section{The model and quantum measures}
The scaled dimensionless two spin $\frac{1}{2}$ Heisenberg $XYZ$ Hamiltonian is given as
\begin{equation}
   H = \frac{J}{2}[(1+\gamma)\sigma_{x}^{1}\sigma_{x}^{2}+(1-\gamma)\sigma_{y}^{1}\sigma_{y}^{2}]+  \frac{1}{2}[J_{z}\sigma_{z}^{1}\sigma_{z}^{2}+B(\sigma_{z}^{1}+\sigma_{z}^{2})] \label{Hamiltonian}
\end{equation}
where $\sigma_{k} $ are the pauli spin matrices, $\gamma = (J_{x}-J_{y})/(J_{x}+J_{y})$ is the anisotropy in $XY$ plane with $J_{k}$ being the interaction strength in respective spin components, $B$ is  the strength of magnetic field. Here the energy is scaled such that $k_{B}T = 1$ where $k_{B} $ is the Boltzmann constant  and $T$ is the equilibrium temperature.
The thermal state of the above Hamiltonian is given by $\rho = e^{- H}/\mathcal{Z}$, where $\mathcal{Z} = \text{tr} (e^{- H})$  is the partition function, and its matrix form in computational basis is obtained as

 \begin{equation}
 \rho = \frac{1}{\mathcal{Z}}
 \begin{pmatrix}
 \mu_{-} & 0 & 0 & \kappa\\
 0 & \nu & \epsilon & 0\\
 0 & \epsilon & \nu & 0\\
 \kappa & 0 & 0 & \mu_{+}\label{Thermal}
 \end{pmatrix}. 
 \end{equation}
 Taking $\eta=\sqrt{B^2+(\gamma J)^2}$, the matrix elements are $\mu_{\pm}=e^{-\frac{J_{z}}{2}}(\cosh \eta ~{\pm}~\frac{B}{\eta}\sinh \eta)$, $\kappa=-\frac{\gamma J}{\eta}e^{\frac{-J_{z}}{2}}\sinh\,\eta$, $\nu = e^{\frac{J_{z}}{2}}\cosh J $,  $\epsilon=-e^{\frac{J_{z}}{2}}\sinh J$ and  $\mathcal{Z}=2(e^{-\frac{J_{z}}{2}}\cosh\eta+e^{\frac{J_{z}}{2}}\cosh J)$.
In order to quantify the correlation between subsystems of bipartite spin states in presence of environment, it is useful to define a set of quantities of our interest as mentioned below.

{\it Concurrence}
\\The entanglement between subsystems of a bipartite state $\rho$ is measured using concurrence \cite{wotters97} as
\begin{equation}
  C(\rho)=\text{max}\lbrace0,\sqrt{\lambda_1}-\sqrt{\lambda_2}-\sqrt{\lambda_3}-\sqrt{\lambda_4}\rbrace 
\end{equation}
where $\lambda_i $ are eigenvalues of matrix $\rho\tilde{\rho}$ arranged in decreasing order and $\tilde{\rho}=(\sigma_y\otimes\sigma_y)\rho^{*}(\sigma_y\otimes\sigma_y)$  is spin flipped matrix. The concurrence lies between $0$ and $1$, such that minimum and maximum values correspond to separable (unentangled) and maximally entangled states respectively.

{\it Measurement-Induced Nonlocality (MIN)}
\\It is a correlation measure in the geometric perspective to capture nonlocal effect on quantum state due to local projective measurements. This quantity in some sense is dual to geometric quantum discord \cite{geo_discord}, and is defined as 
\begin{equation}
N_2(\rho):=~ ^{\text{max}}_{\Pi^a}\Vert\rho-\Pi^a(\rho)\Vert^2
\end{equation}
where $\Vert \mathcal{A} \Vert = \sqrt{\text{tr}(\mathcal{A}^{\dagger}\mathcal{A})}$ is the Hilbert-Schmidt norm of an operator $\mathcal{A}$. Here the maximum is taken over all possible von Neumann projective measurements $\Pi^a=\lbrace\Pi^a_k\rbrace = \lbrace\vert k \rangle \langle k \vert \rbrace ~\text{and}\,\Pi^a(\rho) = \sum_k(\Pi_k^a \otimes \mathbb{I}^b)\rho(\Pi_k^a \otimes \mathbb{I}^b)$.

{\it Trace MIN (T-MIN)}
\\Due to easy computation and experimental realization \cite{exp_real}, much attention has been paid on MIN in recent years. However, this quantity is not a bonafide measure of quantum correlation as it suffers from local ancilla problem \cite{piani}. One alternate form of MIN based on trace distance \cite{tmin}, namely trace MIN (T-MIN) resolves this problem. It is defined as
\begin{equation}
N_1(\rho):= ~^{\text{max}}_{\Pi^a}\vert\rho-\Pi^a(\rho)\vert_1
\end{equation} 
where $\vert \mathcal{A} \vert_1 = \text{tr}\sqrt{\mathcal{A}^{\dagger}\mathcal{A}}$ is the trace norm of operator $\mathcal{A}$. Here also the maximum is taken over all possible von Neumann projective measurements.

{\it Fidelity MIN (F-MIN) }
\\Since fidelity itself is not a metric, any monotonically decreasing function of fidelity defines a valid distance measure. Defining MIN based on fidelity induced metric \cite{fmin} as 
\begin{equation}
N_{\mathcal{F}}(\rho)=~1-~^{\text{min}}_{\Pi^a}\mathcal{F}(\rho,\Pi^a(\rho))
\end{equation}
where $ \mathcal{F}$ is the fidelity between the states $\rho$ and $\rho' $ defined as \cite{fidelity}
\begin{equation}
 \mathcal{F}(\rho,\rho') =\frac{(\text{tr}(\rho\,\rho'))^2}{\text{tr}(\rho)^2 \text{tr}(\rho')^2}. \nonumber
\end{equation}
Here the minimum is taken over all possible projective measurements. This quantity also resolves the local ancilla problem. We shall note that all the three forms of MIN lie between 0 and 0.5, such that minimum and maximum values correspond to uncorrelated and maximally correlated states respectively.
\section{Quantum Channel and correlation}
In this section we investigate the role of environment modelled by quantum channel on the two spin $\frac{1}{2}$ system described by the Hamiltonin (\ref{Hamiltonian}). Influence of environment on the initial thermal state $\rho$ of the system can be described by positive trace preserving operation as
\begin{equation}
\rho' \equiv \varepsilon (\rho) = \sum_{i,j}A_{ij}\rho A_{ij}^\dagger \label{dyn}
\end{equation}
where $A_{ij} = A_i \otimes A_j$, are two qubit Kraus operators satisfying the completeness relation $\sum_{ij} A_{ij}{A^\dagger_{ij}} = \mathbb{I}$.
 Here we shall note that the initial state of the system is given by eq.(\ref{Thermal}), which belongs to the family of $X$-state. Properties of $X$-state are well known and can be found in \cite{vinod}, and references therein. Here we observe that the evolved state under quantum channel retains its $X$-form as 
\begin{equation}
 \rho'= \begin{pmatrix}
    \rho_{11} & 0 & 0 & \rho_{14}\\
    0 & \rho_{22} & \rho_{23} & 0\\
    0 & \rho_{32} & \rho_{33} & 0\\
    \rho_{41} & 0 & 0 & \rho_{44}
  \end{pmatrix}\label{rt}
\end{equation}
where the elements are real.
The concurrence, MIN, T-MIN and F-MIN for the above state are computed as
\begin{equation}
C(\rho')=2\,\text{max}\left\lbrace 0,\vert\rho_{14}\vert-\sqrt{\rho_{22}\,\rho_{33}},\vert\rho_{23}\vert-\sqrt{\rho_{11}\,\rho_{44}} \right\rbrace \, , 
\label{concurence}
\end{equation} 
\begin{equation}
N_2(\rho') = 2\,(\rho_{14}^2 + \rho_{23}^2),
\end{equation}
\begin{equation}
N_{1}(\rho') = \vert\rho_{14}\vert + \vert\rho_{23}\vert,
\end{equation}
\begin{equation}
N_{\mathcal{F}}(\rho') = \frac{2(\rho_{14}^2 + \rho_{23}^2)}{2(\rho_{14}^2 + \rho_{23}^2)+(\rho_{11}^2 + \rho_{22}^2 + \rho_{33}^2 + \rho_{44}^2)} \,.
\end{equation}
It is clear from the above results that all the four quantities vanish identically if $\rho_{14} = \rho_{23} = 0$, which corresponds to uncorrelated state. We also note that, $C(\rho) = 0$ if $| \rho_{14}|  \leq  \sqrt{\rho_{22}\rho_{33}}$ and $\lvert \rho_{23} \rvert \leq \sqrt{\rho_{11}\rho_{44}}$. In other words, the state influenced by the quantum channel can be unentangled with non-zero MINs. In this sense, concurrence and MINs quantify different aspects of non-locality of quantum state. 

It is worth recognising that while the off-diagonal elements of density matrix arise from the superposition of states and thus signify the quantum signature, diagonal elements signify the statistical mixture of quantum ensemble. We observe from the above results that, while MIN and T-MIN are obtained from off-diagonal elements of the density matrix, F-MIN is obtained from both off-diagonal and diagonal elements. Hence F-MIN could be more useful correlation measure than the other two MINs to classify the bipartite states. Since all the above MINs range from 0 to 0.5, the factor 2 in the concurrence is discarded for better numerical comparison. In what follows, we compute the state of the system under the influence of two quantum channels.
\subsection{Generalized Amplitude Damping Channel}
Let us consider a single qubit whose ground and excited states are $\lvert g \rangle$ and $\lvert e \rangle$ respectively. If the qubit interacts with an environment such that it decays from excited state to the ground state, we say that a photon is emitted by the system. Such an environment is modelled by amplitude damping (AD) channel. However, in general the interaction is such that energy is exchanged between the qubit and environment in both ways, that is in the form of emission and absorption of photon. If $p$ is the probability of emission, then $(1-p)$ is the probability of absorption. In such a process, the probability $p$ is proportional to the temperature difference between the quantum system and the environment. That is, higher the temperature of the system than the environment, greater is the probability of emission process. Such an interaction can be modelled by generalized amplitude damping (GAD) channel whose Kraus operators are 

\begin{equation}
  A_0 = \sqrt{p}
  \begin{pmatrix}
    1 & 0\\
    0 & \sqrt{1-\lambda}
  \end{pmatrix}
  ,~ A_1 = \sqrt{p}
  \begin{pmatrix}
    0 & \sqrt{\lambda}\\
    0 & 0
  \end{pmatrix}
  $$
  $$
  A_2 = \sqrt{1-p}
  \begin{pmatrix}
    \sqrt{1-\lambda} & 0\\
    0 & 1
  \end{pmatrix}
 ,~ A_3 = \sqrt{1-p}
  \begin{pmatrix}
    0 & 0\\
    \sqrt{\lambda} & 0
  \end{pmatrix}
\end{equation}
where $\lambda =1-e^{-\Gamma t}$, with $\Gamma$ being the spontaneous decay rate. Here $\lambda$ is sometimes referred as decoherence parameter or the scaled time such that $t \in [0,\infty] $ is mapped on to $\lambda \in [0,1]$. It is clear that $p = 1$ corresponds to the AD channel.

Defining a single qubit state as
\begin{equation}
\sigma_{\infty} =
  \begin{pmatrix}
    p & 0\\
    0 & 1-p
  \end{pmatrix}
\end{equation}
it is straight forward to check that $\varepsilon'(\sigma_{\infty}) = \sigma_{\infty}$ where $\varepsilon'(\sigma) =\sum_k A_k \sigma A^{\dagger}_k $.
Here $A_k$ are the Kraus operators on single qubit satisfying the completeness relation $\sum_k A_k A^{\dagger}_k = \mathbb{I}$.
In other words, $\sigma_{\infty}$ is a steady state of single qubit under this channel. Constructing a two qubit state
\begin{equation}
  \rho_s = \sigma_{\infty} \otimes \sigma_{\infty} = 
  \begin{pmatrix}
    p^2 &  0      &   0      &     0 \\
    0   &  p(1-p) &   0      &     0 \\
    0   &  0      &   p(1-p) &     0 \\
    0   &  0      &   0      &     (1-p)^2
  \end{pmatrix}\label{stationary}
\end{equation}
we can verify that $\varepsilon(\rho_s) = \rho_s$. This means that $\rho_s$ is a steady state of two qubit under this channel. In other words, a two qubit system in some arbitrary initial state will evolve into this state $\rho_s$ asymptotically, under the influence of GAD quantum channel. We also note that the state $\rho_s$ is diagonal and so is an uncorrelated state. Here we observe that $p$ characterizes probability distribution of the steady state in computational basis. 
If the probabilities of emission and absorption are same $(p = 0.5)$, then the steady state is maximmally mixed i.e., $\rho_s = \mathbb{I}/4$.  

If we consider the initial state as the thermal state (\ref{Thermal}), then the evolved state under this channel is computed as follows:
\begin{align} \nonumber
\rho_{14} &= \rho_{41} = \kappa (1-\lambda) \\ \nonumber
\rho_{23} &= \rho_{32} = \epsilon (1-\lambda)  \\ \nonumber
\rho_{11} &= \mu_-(1+q\lambda)^2 + p\lambda[2\nu r + (2\nu +\mu_+) p \lambda ] \\ \nonumber
\rho_{22} &= \rho_{33} = \nu(r - pq\lambda) - \lambda[\mu_-q(1+q\lambda) + p\{\nu\,q\,\lambda  +\mu_+(p\lambda-1)\}] \\ \label{gag_elem}
\rho_{44} &= \mu_+(p\lambda-1)^2 + q\,\lambda[\mu_-q\lambda + 2\nu (p\,\lambda-1)]  \\ \nonumber
\end{align} 
where $q = p-1$. From this, we immediately observe that the off-diagonal elements are independent of $p$ and they vanish at $\lambda = 1$, the asymptotic limit. That is, the initial thermal state of the system eventually evolves into uncorrelated state. The system parameters and $p$ only alter the distribution of mixture in the asymptotic limit.
\subsection{Hybrid Channel}

This channel is derived from three single qubit operations namely Bit-Flip (BF), Phase-Flip (PF) and Bit-Phase-Flip (BFP). The Kraus operators associated to these operations are listed in Table \ref{kraus_tab}. From the table, the operation BF is understood as application of $\sigma_x$ (NOT) to a single qubit with probability $p$ and application of $\mathbb{I}$ (identity) with probability $(1-p)$. Similarly, BPF and PF operations are understood with $ \sigma_y$ and $ \sigma_z$ respectively. The hybrid channel on two qubit is thus constructed  from applying the above three operations on two qubits with weight factors $ \alpha, \beta, $ and $\delta$ such that $\alpha + \beta + \delta = 1$. Such a channel is represented as 
\begin{equation}
  \varepsilon(\rho) = \alpha~\varepsilon_{BF}(\rho) + \beta~\varepsilon_{PF}(\rho) +\delta~\varepsilon_{BPF}(\rho). \label{sum}
\end{equation}\label{sum}
\begin{table}
  \centering
  \begin{tabular}{c c c}
  \hline\hline
   Operation &     $A_0$   &  $A_1$ \\
   \hline
   BF &       $ \sqrt{1-p}\,\mathbb{I}$  &     $\sqrt{p}\, \sigma_x$\\
   BPF &      $ \sqrt{1-p}\,\mathbb{I}$  &    $\sqrt{p}\, \sigma_y$\\
   PF &       $ \sqrt{1-p}\,\mathbb{I}$  &     $\sqrt{p}\, \sigma_z$\\
   \hline
  \end{tabular}
  \caption{Kraus operators}
  \label{kraus_tab}
\end{table}
 If we allow the initial state (\ref{Thermal}) to evolve under this channel, elements of the evolved state are given by:
\begin{align} \nonumber
\rho_{14} &= \rho_{41} = \frac{1}{2} [-\epsilon m p (\alpha - \beta) +  \kappa \psi]\\ \nonumber
\rho_{23} &= \rho_{32} = \frac{1}{2} [-\kappa m p (\alpha - \beta) +  \epsilon \psi]\\ \nonumber
\rho_{11} &= \frac{1}{4}[p(\alpha + \beta) \{\mu_+\, p -2 \nu m\} +\mu_-\chi]\\ \nonumber
\rho_{22} &= \rho_{33} = \frac{1}{4} [p(\alpha + \beta) \{\nu\, p -(\mu_+ + \mu_-)m\} +\nu \chi]\\
\rho_{44} &= \frac{1}{4}[p(\alpha + \beta) \{\mu_-\, p -2 \nu m\} +\mu_+\chi]
\end{align} 
where $m = p-2$, $n = p-1$, $l = 2-2p+p^2$, $\chi = (\alpha + \beta)m^2 + 4 \delta $ and $\psi = (\alpha + \beta)l + 2 n^2 \delta$.
\section{Discussion}
In what follows we analyse in detail the correlation between two spin states in presence of environment using various forms of MINs. The correlation quantified using MINs are then compared with entanglement between the spins measured by concurrence. 

\begin{figure}[h!]
 \includegraphics[width=0.45\linewidth]{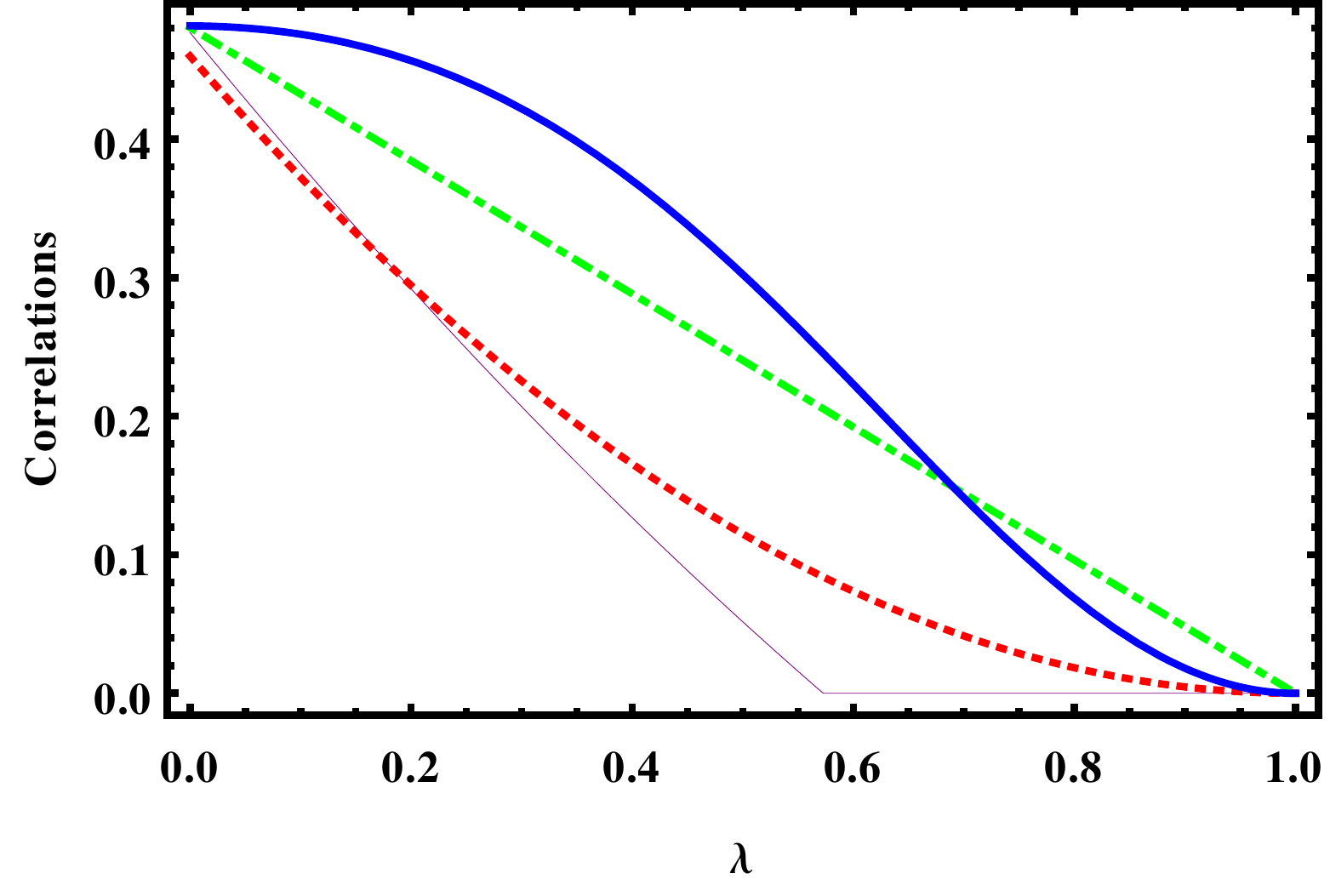}
 \includegraphics[width=0.45\linewidth]{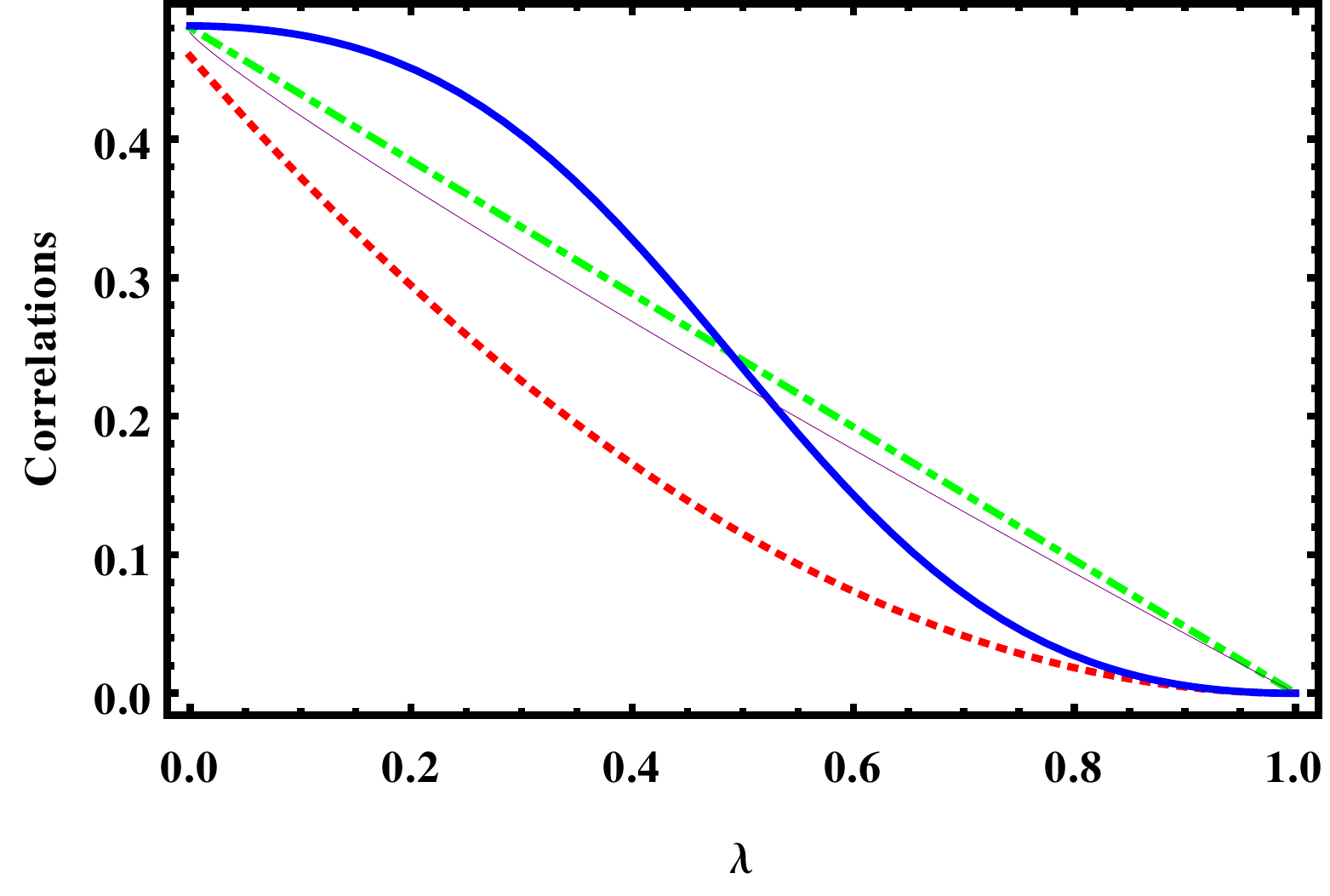}
 \caption{(Color online)  MIN (dotted), T-MIN (dash-dotted), F-MIN (thick) and concurrence (thin)  as a function of decoherence factor $\lambda$ for $p = 0.5 $ (left), $p = 0,1$ (right) with $B = 0, \gamma = 0.1, J = 2, J_z = 2.$}
 \label{gad_deco} 
\end{figure}

\begin{figure}[h!]
 \includegraphics[width=0.45\linewidth]{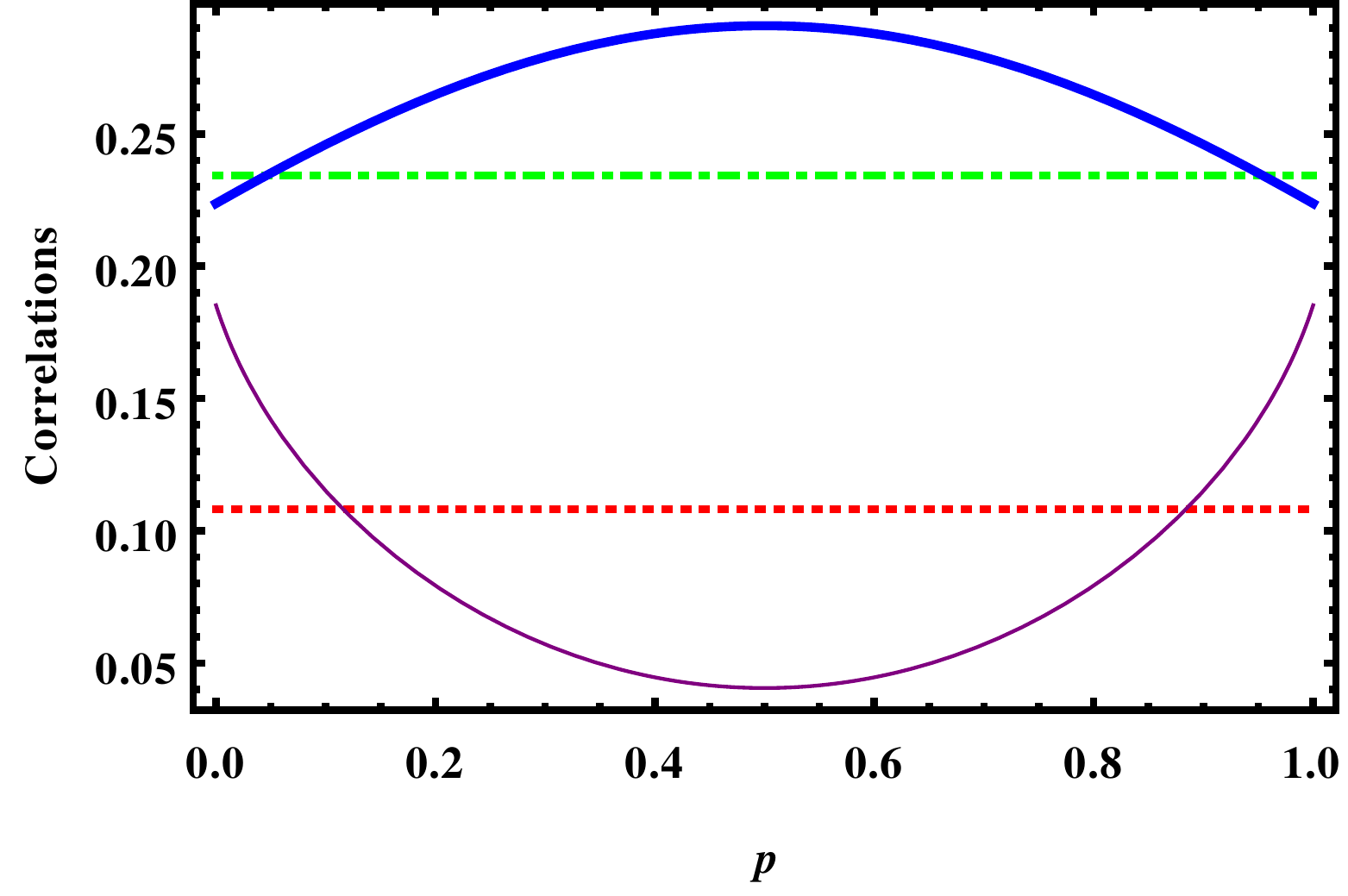}
 \includegraphics[width=0.45\linewidth]{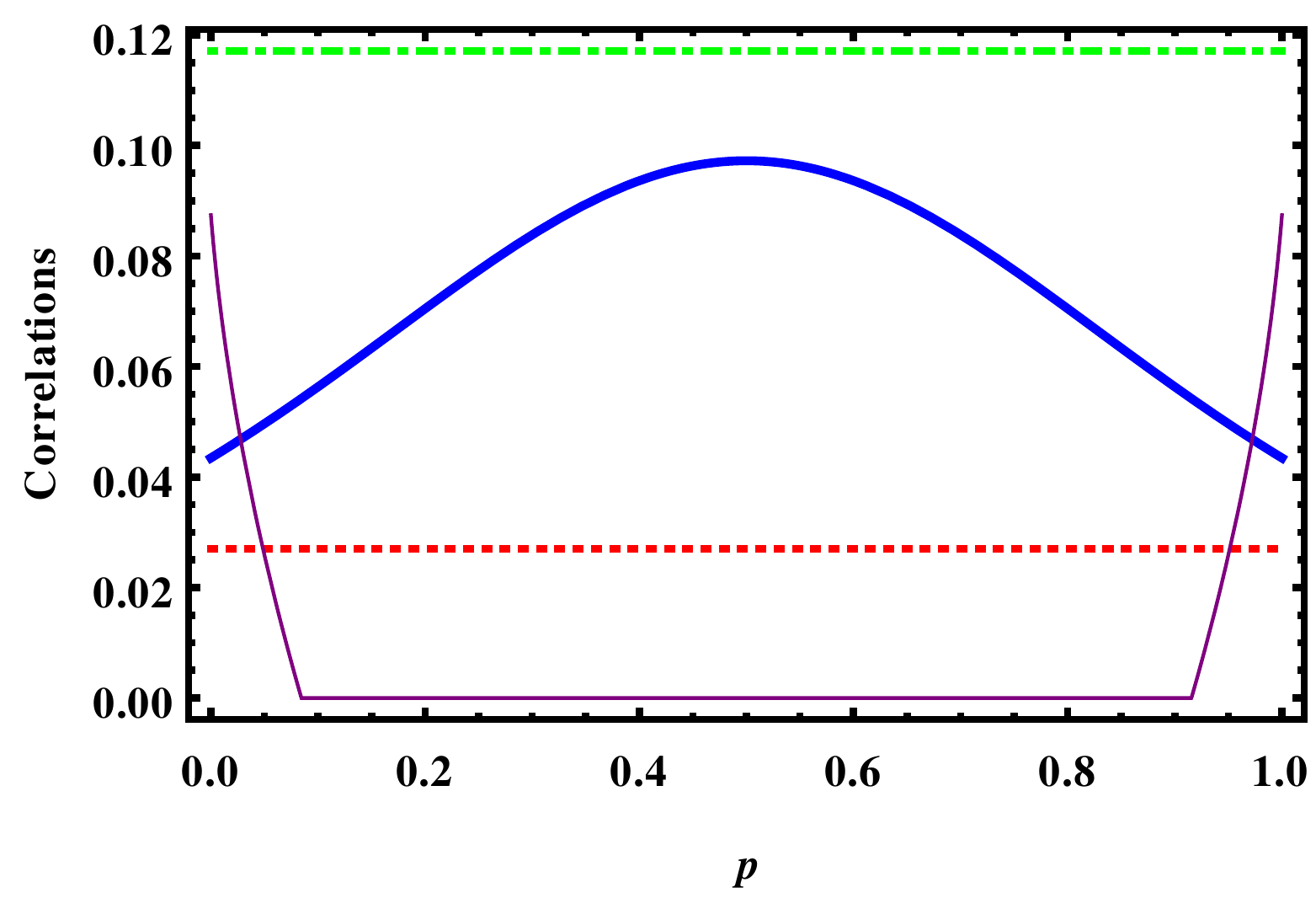}
 \caption{(Color online) MIN (dotted), T-MIN (dash-dotted), F-MIN (thick) and concurrence as a function of $p$  for $B = 0, \gamma = 0.1, J = 2, J_z = 2 $ with $\lambda = 0.5$ (left) and $\lambda = 0.75 $ (right).}
 \label{gad_deco_p}
\end{figure}

To begin with, we look at the time evolution of MINs and concurrence as a function of decoherence factor 
$\lambda$ for the environment modelled by the GAD channel. It is observed from eq.(\ref{gag_elem}) that since the elements $\rho_{14}$ and $\rho_{23}$ of the evolved state are independent of $p$ (probability of emission), the quantum correlation measured by F-MIN is dependant on $p$, unlike MIN and T-MIN. This implies that F-MIN is a good measure of quantum correlation in such process. We have plotted the time evolution of MINs and concurrence for $p = 0.5, 1$, as shown in Fig. \ref{gad_deco}. Here we observe that the correlation quantified by MIN, T-MIN and F-MIN are decreasing with $\lambda$ or time $t$, and they vanish only in the asymptotic limit $(\lambda = 1)$, wherein the system reaches the corresponding steady state $ \rho_s$ as mentioned earlier. On the other hand, the concurrence vanishes for $\lambda \geq \lambda_c$ where $\lambda_c$ is some critical value which depends on $p$ for given system parameters. This is known as the sudden death of entanglement \cite{sudn_death}. That is the two spins are unentangled in some finite time, even though the correlation quantified by MINs do not vanish. From our numerical analysis, we also found that $\lambda_c$ is minimum for $p = 0.5$ where the system has an equal probability of emission and absorption, leading to an early sudden death. 
In Fig. \ref{gad_deco_p} we look at the variation of MINs and concurrence with $p$. Though MIN and T-MIN are independent of $p$ (as mentioned earlier), F-MIN and concurrence are not. For $\lambda = 0.5$, we observe that F-MIN is maximum at $p = 0.5$ wherein the entanglement is minimum. For $\lambda = 0.75$, while the entanglement is shown to be zero over a wide range of $p$, F-MIN is non-zero throughout with maximum at $p = 0.5$. In other words, the quantum correlation as measured by F-MIN is maximum when absorption and emission are equally probable. These observations also indicate that nonlocality manifested by entanglement is completely different in nature than that is captured by different forms of MIN. Further, it appears from our observation that F-MIN is more useful to classify the non-locality of quantum states than the other two MINs.

\begin{figure}[h!]
\includegraphics[width=0.45\linewidth]{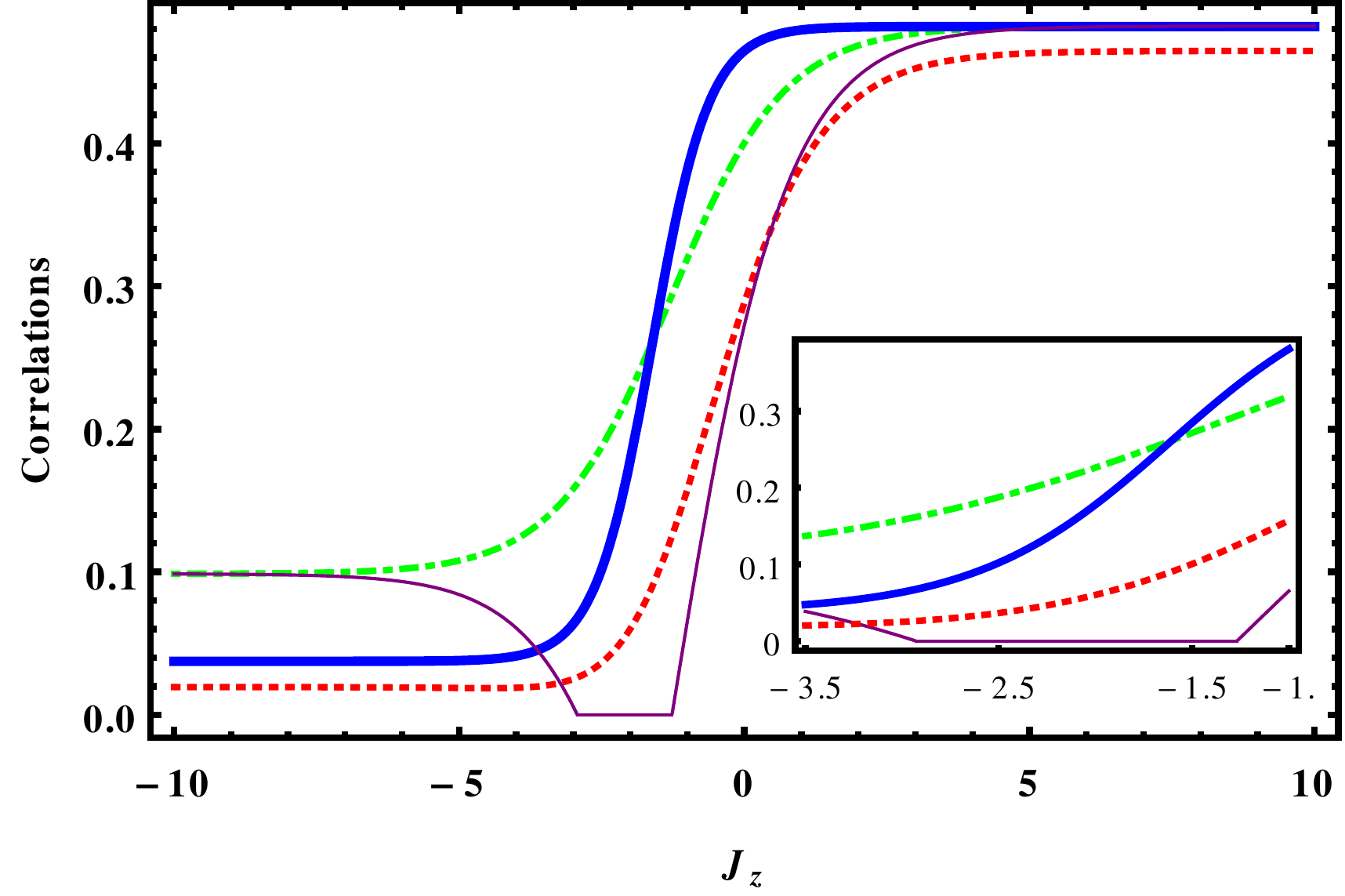}
 \includegraphics[width=0.45\linewidth]{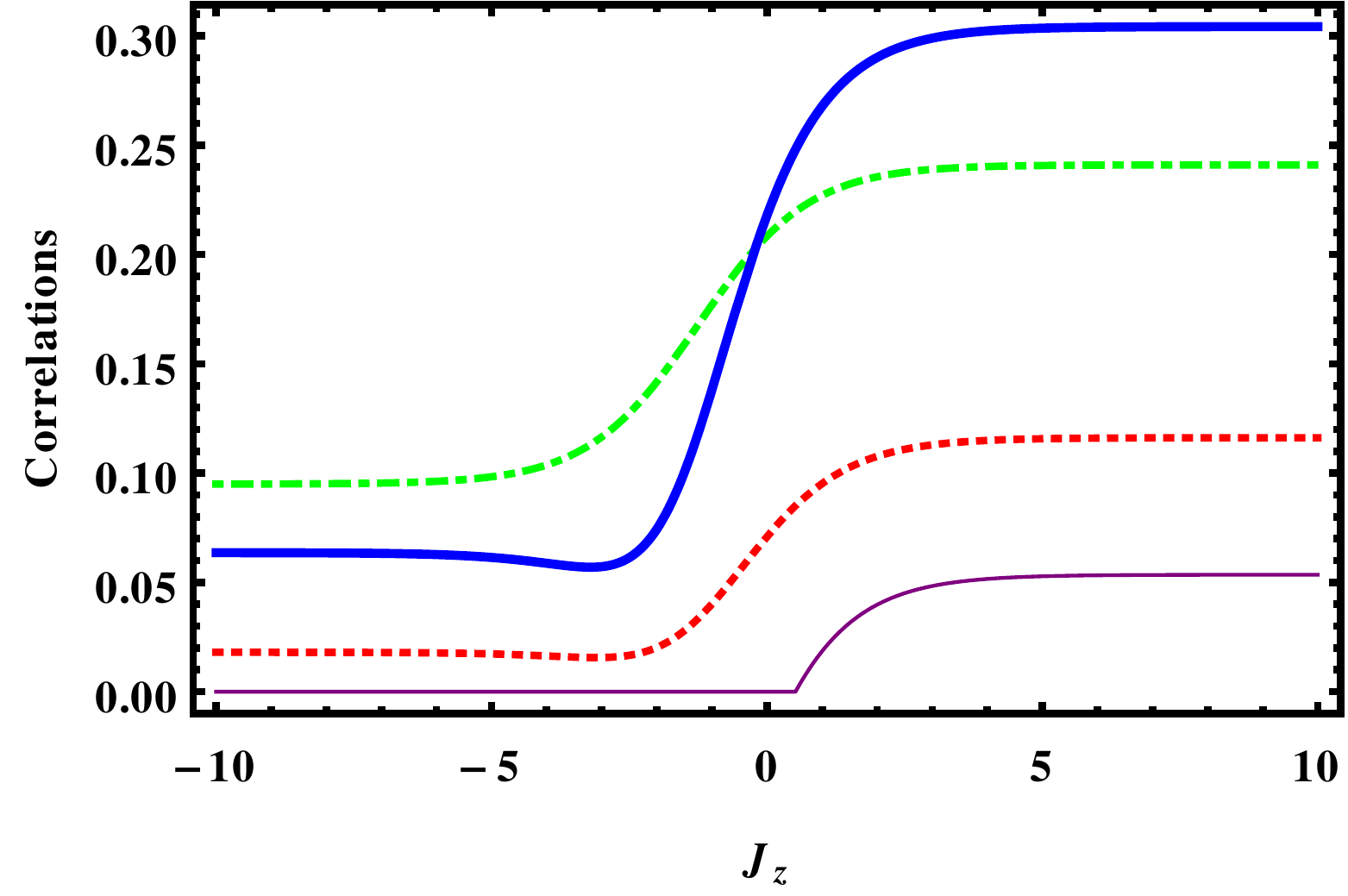}
 \caption{(Color online) MIN (dotted), T-MIN (dash-dotted), F-MIN (thick) and concurrence as a function of $J_z$ for  $B = 0, \gamma = 0.1, J = 2$. (a) without channel and (b) under the influence of GAD channel for $p = 0.5$ and $\lambda = 0.5$.}
 \label{gad_deco_J}
\end{figure}

Now we make a comparison of quantities of our interest for the quantum state without and with channel as shown in Fig. \ref{gad_deco_J}. In the absence of GAD channel, while the entanglement is zero over a range of parameters as predicted earlier \cite{my}, the MINs do not vanish anywhere. On the other hand, the channel significantly affects entanglement so that the spins are unentangled over a large range of parameters.
Though the MINs are reduced due to the channel, they do not vanish unlike the entanglement.

\begin{figure}[h!]
 \includegraphics[width=0.45\linewidth]{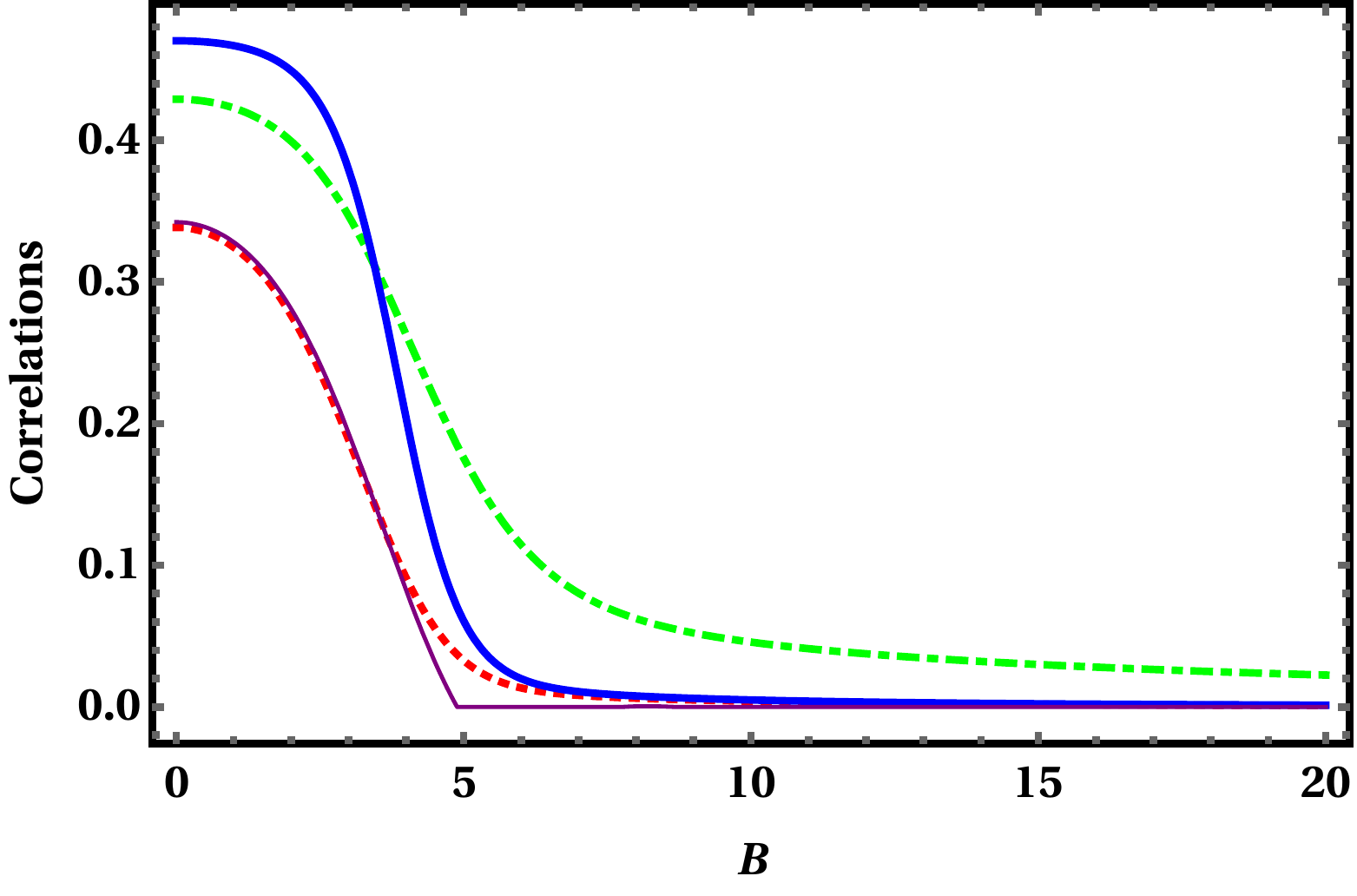}
 \includegraphics[width=0.45\linewidth]{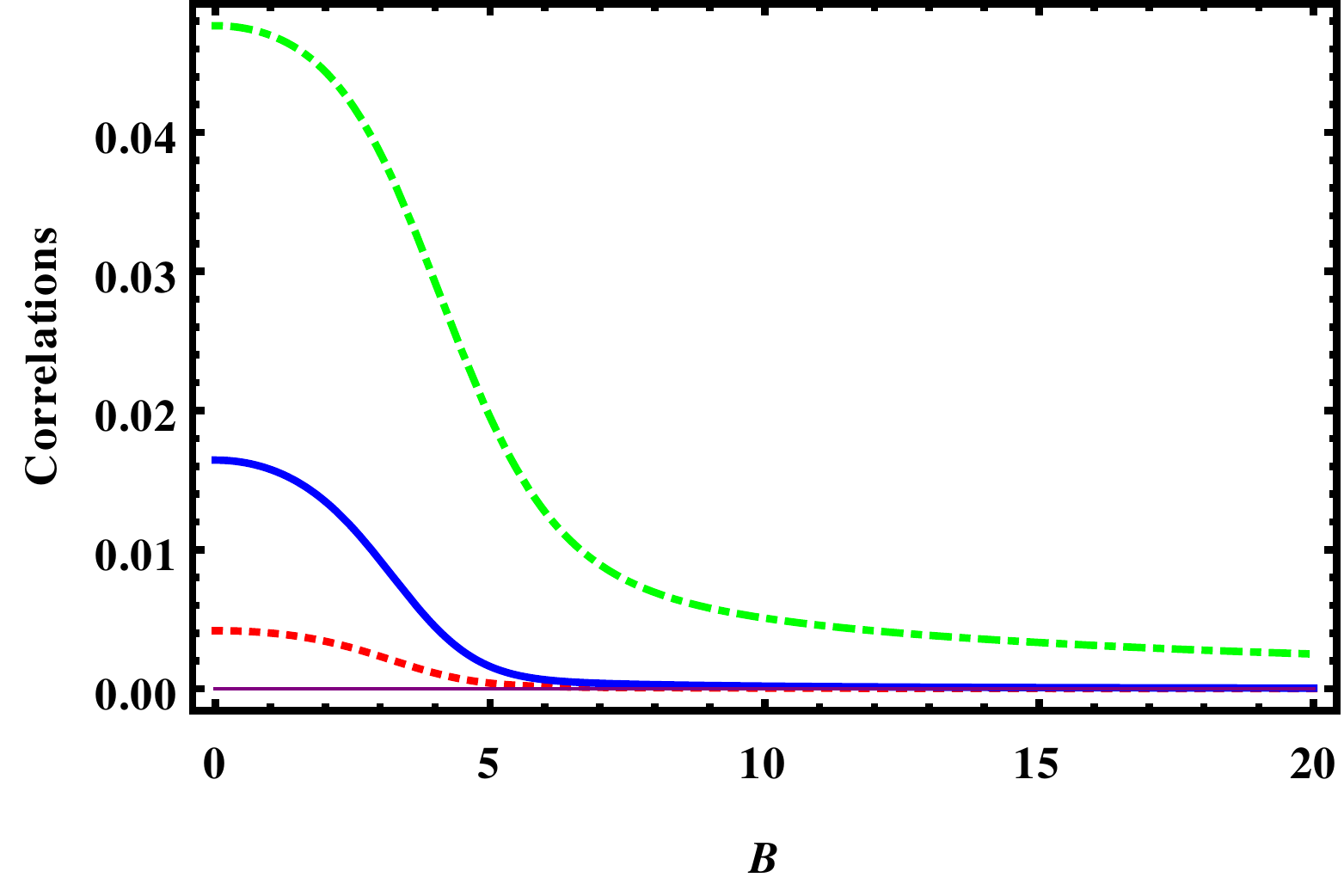}
 \caption{(Color online) MIN (dotted), T-MIN (dash-dotted), F-MIN (thick) and concurrence for $ \gamma = 0.1, J = 2, J_z = 2$ with $ p = 0.5$ at $\lambda = 0.1$ (left)~and ~$\lambda = 0.9$ (right).}
 \label{gad_deco_B}
\end{figure}

To observe the role of external magnetic field on the correlation measure, we plot our results as shown in Fig. \ref{gad_deco_B}. Here we observe that all the quantities are shown to decrease with the increase of magnetic field $B$. As time progresses (for large $\lambda$), the influence of channel is also visible with significant reduction in the correlation.

\begin{figure}
  \includegraphics[width=0.45\linewidth]{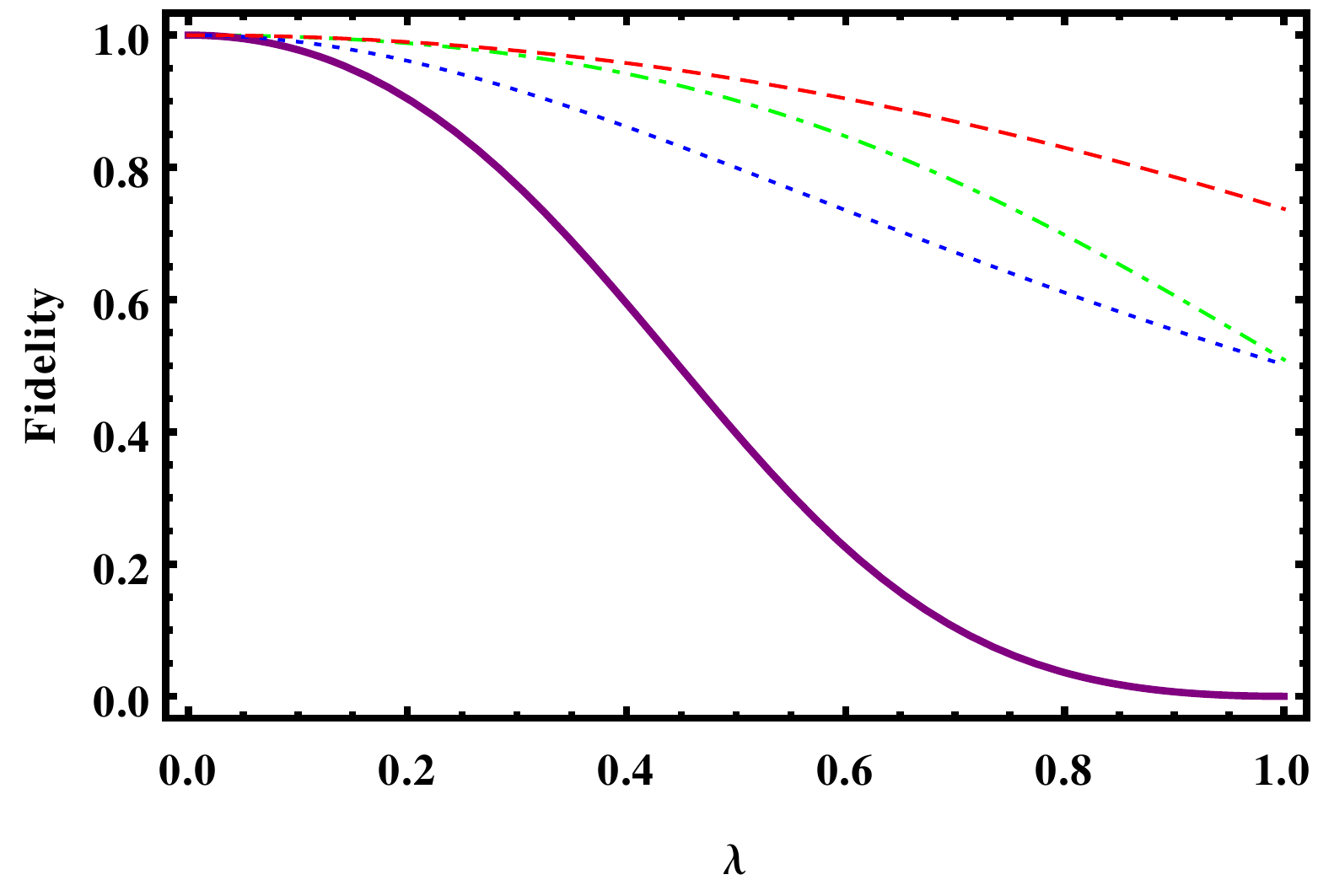}
  \includegraphics[width=0.45\linewidth]{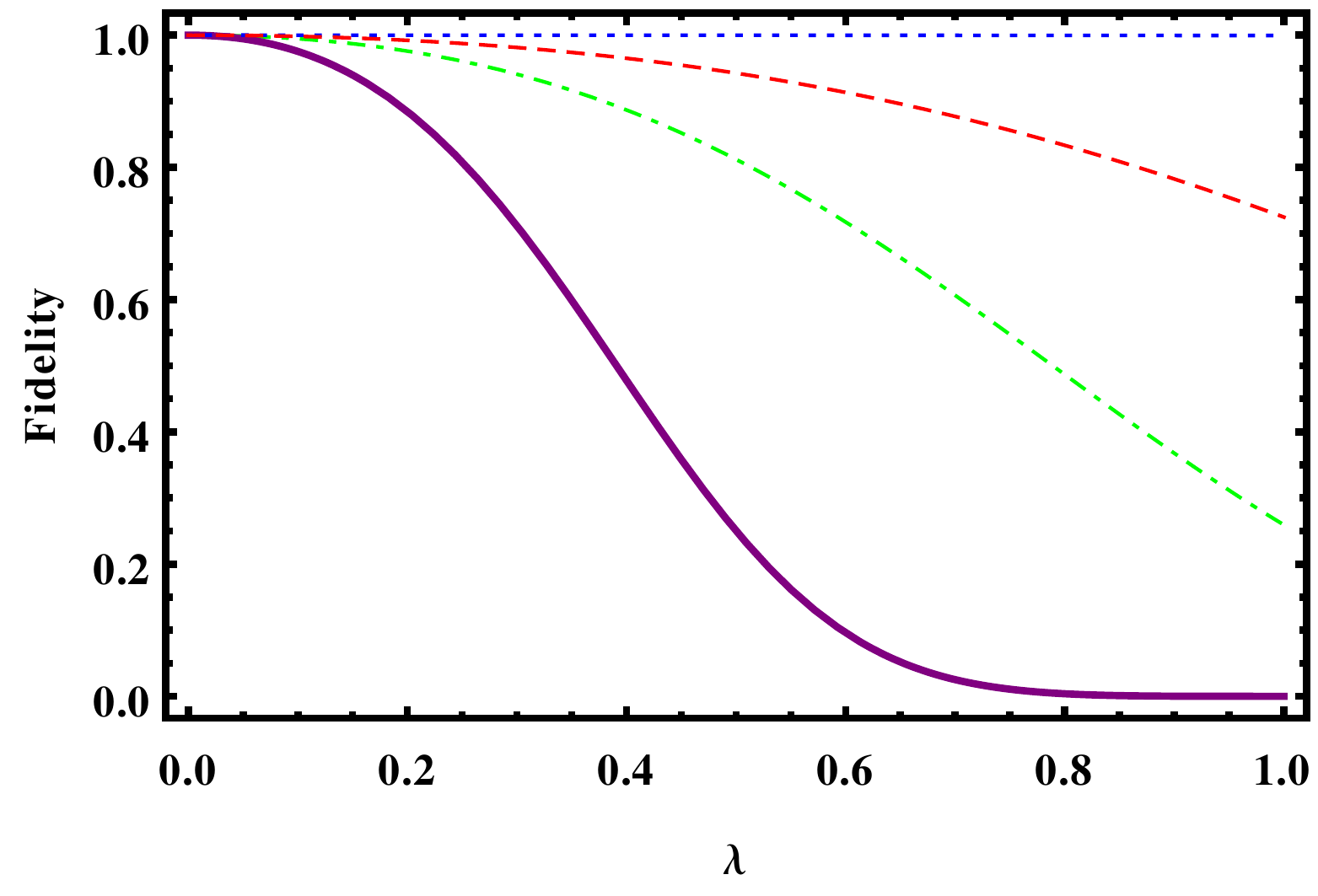}
 \caption{(Color online) Fidelity of GAD channel for p = 1 (thick), 0.5 (dot-dashed), 0.3 (dashed), 0  (dotted) with $ \gamma = 0.1, J = 2, J_z = 2,$ for $ B = 4$ (left) and $B = 8 $ (right).}
 \label{fidility}
 \end{figure}

An alternate quantity of interest in studying the dynamics of quantum state is the overlap between initial and the evolved state, which may be quantified using fidelity $ \mathcal{F}(\rho,\rho')$.
We shall note that $0 \leq \mathcal{F}(\rho,\rho') \leq 1$ with maximum fidelity for $\rho = \rho'$, and minimum if the states do not overlap. Since we are interested in computing fidelity between the initial thermal state and the state evolved under the quantum channel, 
this fidelity can also be thought of as characteristic of the given channel, and may be referred as channel fidelity. Here we plot fidelity for GAD channel with respect to the decoherence parameter $\lambda$ as shown in Fig. \ref{fidility} for high magnetic field $B$. In general, the channel fidelity decreases from one, and the rate of decrease crucially depends on the probability $p$. In the asymptotic limit, the fidelity quantifies the overlap between the initial thermal state (\ref{Thermal}) and the steady state $\rho_s$ as defined by (\ref{stationary}). 

We also observe that the fidelity settles down to zero for $p = 1$, wherein the absorption is prohibited. In other words, in presence of quantum channel the system settles down to the pure sate $\lvert 00\rangle \langle 00 \rvert$, which is orthogonal to the initial state. On the other hand, the fidelity remains one for $p = 0$ i.e., if the emission of energy from system to environment is prohibited. It implies that the initial state of the system is pure ie., $\rho = \lvert 11\rangle \langle 11 \rvert$ for sufficiently larger magnetic field, and is unaffected by the influence of quantum channel as long as energy is not dissipated in the form of emission. In this way, a controlled environment can facilitate for preparing the system in specific pure state. For all other values of $p$, the system settles down to mixed uncorrelated state as defined in eq. (\ref{stationary}).

\begin{figure}[h!]
  \includegraphics[width=0.45\linewidth]{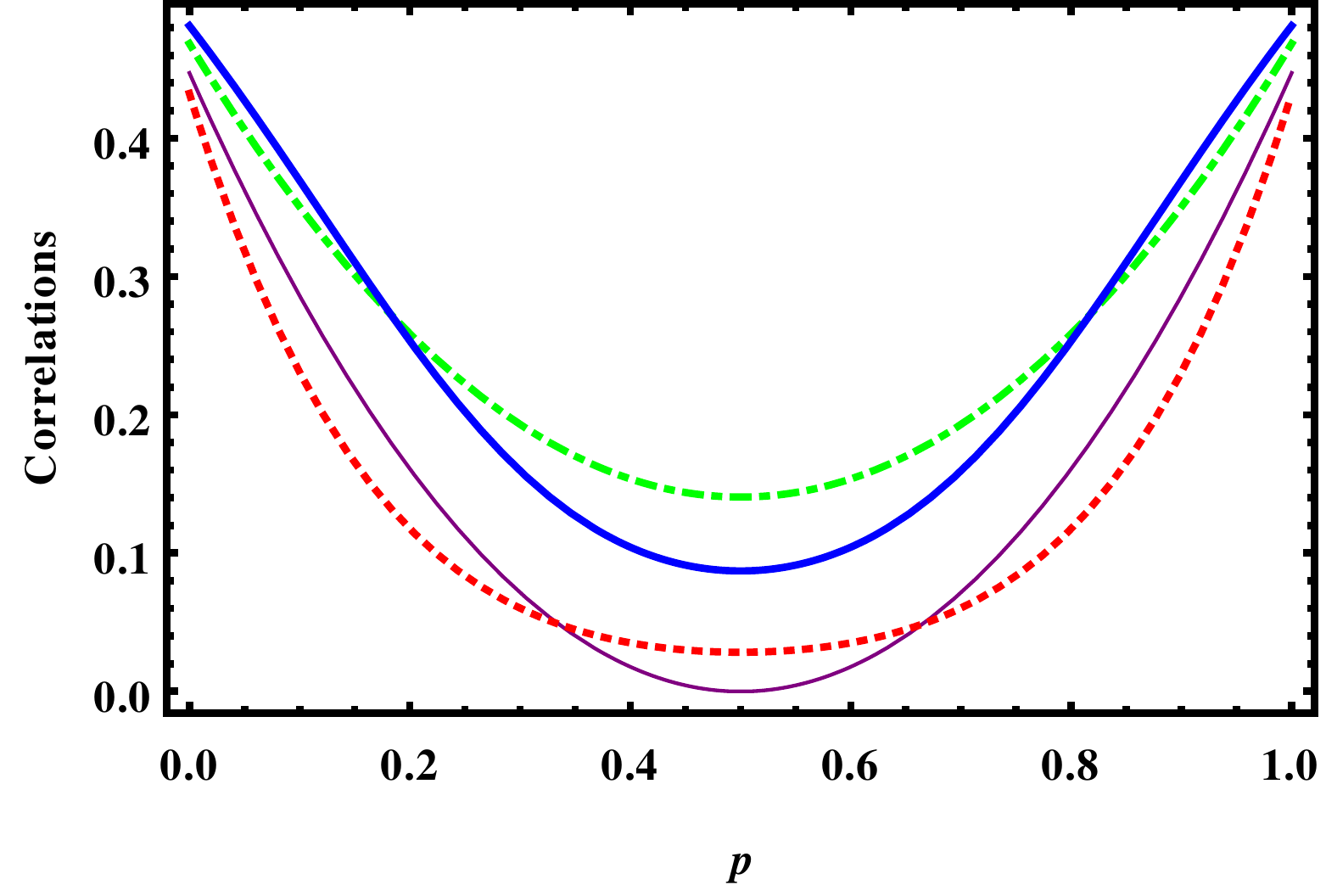} 
  \includegraphics[width=0.45\linewidth]{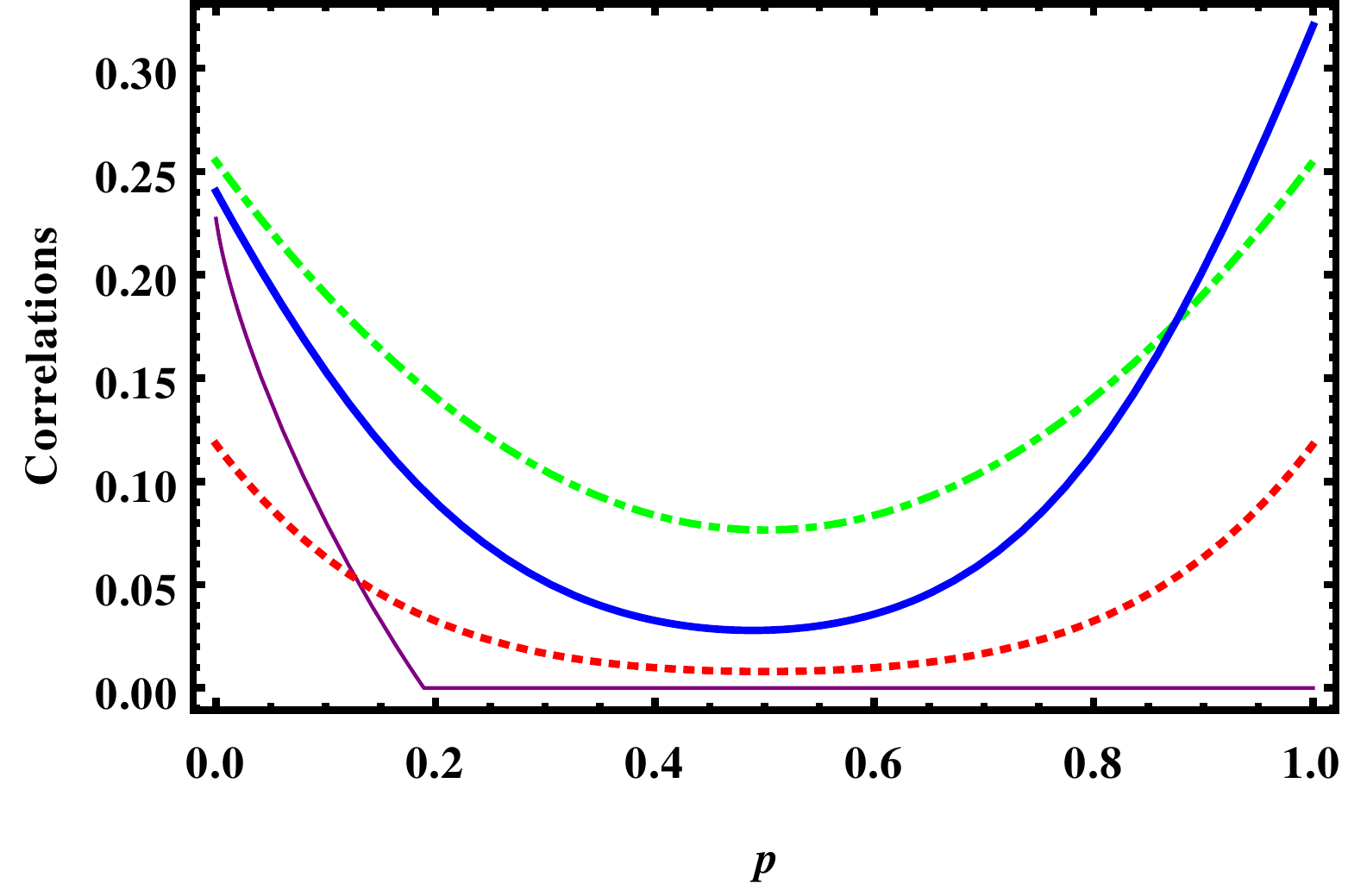} 
 \caption{(Color online) MIN (dotted), T-MIN (dash-dotted), F-MIN (thick) and concurrence as a function of $p$ for with $B = 0$, $B= 4$, with $\gamma = 0.1, J = 2, J_z = 2$. Here we take ($\alpha , \beta , \delta) =  (0.3, 0.2, 0.5)$.}
 \label{hybrid_p}
\end{figure}

Finally we compare the entanglement of two-spin state with the three forms of MIN in presence of hybrid channel, and few typical observations are shown in Fig. \ref{hybrid_p}. In the absence of magnetic field $B$, the entanglement and MINs exhibit similar behaviour and are symmetric about the flipping probability $p = 0.5$. On the other hand, we observe that the magnetic field $B$ destroys the entanglement between the spins for a wide range of $p$, wherein the MINs are non-zero - showing that the spins states possess quantum correlation without being entangled. We also notice that F-MIN is not symmetric about $p = 0.5$, unlike the other companion MINs, indicating that F-MIN is more sensitive to the probability of flipping operations. These observations once again favour that F-MIN is a more useful quantity in characterizing the quantum signature of states than the other MINs. 
\section{Conclusion}
In this article, we have studied the role of environment modelled by GAD and hybrid channels on entanglement and MINs for Heisenberg $XYZ$ spin system with magnetic field. A detailed analysis of spontaneous emission and absorption in $XYZ$ spin in a finite temperature environment is investigated using GAD channel. We have shown that in the asymptotic limit the system settles down to an uncorrelated steady state which depends only on the probability of emission. While quantum correlation between the spins is shown to exist without being entangled, the correlation vanishes only in the asymptotic limit. Our analysis using channel fidelity indicates the possibility of preparing bipartite pure state at high magnetic field, i.e in the Stern-Gerlach type apparatus. We also demonstrate that F-MIN is more useful than MIN and T-MIN to characterize the correlated quantum states.


\begin{thebibliography}{99}
\bibitem{epr}A. Einstein, B. Podolsky and N. Rosen, \emph{Phys. Rev.} \textbf{47} (1935) 777. 
\bibitem{bel}J. S. Bell, \emph{Physics} \textbf{1} (1964) 195.
\bibitem{werner}R. F. Werner, \emph{Phys. Rev. A} \textbf{40} (1989) 4277.
\bibitem{discord}H. Olliver and W. H. Zurek, \emph{Phys. Rev. Lett.} \textbf{88} (2001) 017901.
\bibitem{min}S. Luo and S. Fuo, \emph{Phys. Rev. Lett.} \textbf{106} (2011) 120401.
\bibitem{tmin}M. L. Hu and H. Fan, \emph{New. J. Phys.} \textbf{17} (2015) 033004.
\bibitem{fmin}R. Muthuganesan and R Sankaranarayanan, \emph{Phys. Lett. A} \textbf{381} (2017) 3028.
\bibitem{piani}M. Piani, \emph{Phys. Rev. A} \textbf{86} (2012) 034101.
\bibitem{corltn}M. A. Nielson and I. L. Chuang Quantum Computation and Quantum Information, Cambridge University Press, UK, (2000).
\bibitem{wotters97}W. K. Wootters, \emph{Phys. Rev. Lett.} \textbf{80} (1997) 2245.
\bibitem{geo_discord}B. Dakic, V. Vedral, and C. Brukner, \emph{Phys. Rev. Lett.} \textbf{105} (2010) 190502.
\bibitem{exp_real}D. Girolami and G. Adesso, \emph{Phys. Rev. Lett.} \textbf{108} (2012) 150403.
\bibitem{fidelity}X. Wang, C. Shui and X. X. Yi, \emph{Phys. Lett. A} \textbf{373} (2008) 58.
\bibitem{vinod} N. Vinod, R. Muthuganesan and R. Sankaranarayanan, \emph{arXiv:1708.06632}.
\bibitem{sudn_death}Yu. T, Eberly. J. H, \emph{Science} \textbf{293} (2009) 5914. 
\bibitem{my}Indrajith. V. S, R. Muthuganesan, R. Sankaranarayanan, \emph{Physica A} \textbf{527} (2019) 121325.
\end{thebibliography}
\end{document}